\documentclass[aip,reprint]{revtex4-1}
\usepackage{graphicx}
\usepackage{upgreek}

\draft 

\begin{document}


\title{Evidence for quantized vortex entry in Nb$_3$Sn under the influence of RF fields} 



\author{Daniel Leslie Hall}
\email[]{dlh269@cornell.edu}
\affiliation{Cornell Laboratory for Accelerator-Based Sciences and Education, Ithaca, New York 14853, USA}

\author{Matthias Liepe}
\email[]{mul2@cornell.edu}
\affiliation{Cornell Laboratory for Accelerator-Based Sciences and Education, Ithaca, New York 14853, USA}

\author{Ryan D. Porter}
\affiliation{Cornell Laboratory for Accelerator-Based Sciences and Education, Ithaca, New York 14853, USA}


\date{\today}

\begin{abstract}
Evidence for quantised magnetic vortex entry at a defect into a Type-II superconductor operating in the metastable state above $B_{c1}$ while under the influence of RF electromagnetic fields has been observed in a polycrystalline coating of Nb$_3$Sn. The measurement was made using a 1.3~GHz microwave cavity for use in particle accelerator R\&D and equipped with a temperature mapping system for precision measurements of local heating. At peak RF fields near the quench field of the cavity, discrete jumps in local heating at the quench origin were observed. A number of these jumps were found to be identical in height within measurement error. This suggests entry of a limited number of vortices at a defect. The behaviour observed appears to be consistent with recent theoretical predictions regarding vortex entry at grain boundaries in the material.
\end{abstract}


\maketitle 



The expectation that ideal Type-II superconductors should maintain a flux-free Meissner state when subject to AC electromagnetic fields above the lower critical field $H_{c1}$, up to the superheating field $H_{sh}$ (which can be many times higher than $H_{c1}$) \cite{Matricon1967,Chapman1995,Transtrum2011,Valles2011}, is crucial for the operation and development of future superconducting microwave particle accelerator cavities capable of generating very high energy beams. In AC superconducting applications the entry of magnetic flux during the cycle, whose frequency can range from MHz to many GHz, will result in heat dissipation far in excess of that experienced due to BCS losses, due to the normal conducting vortex cores experiencing drag \cite{PhysRev.140.A1197,nozieres1966} as they move through the superconductor. These losses will result in a temperature rise in the superconductor that often results in a feedback mechanism that further lowers the barrier for flux entry, inevitably taking the superconductor above its critical temperature and resulting in a loss of the superconducting state and, by extension, loss of function of the superconducting device. In the study of superconducting microwave cavities, and in this paper, such an event is referred to as a \emph{quench}. Ensuring stability and reliability to operate in this metastable state is therefore crucial to these applications.

Although weakly Type-II superconductors such as niobium have been demonstrated to reliably operate in the metastable state\cite{Valles2011,PhysRevLett.115.047001}, the higher $\kappa$ superconductors such as the A15 compounds and cuprates, made attractive by their comparatively higher values of transition temperature $T_c$ and superheating field $H_{sh}$, often exhibit limitations that prevent operation far into the metastable state. These can largely be attributed to their shorter coherence length $\xi$ rendering them more sensitive to defects. This is exacerbated by the compound nature of these superconductors, which allows for off-stoichiometric regions, and the more complex fabrication methods, which can introduce chemical impurities and structural imperfections. Use of these high-$T_c$ superconductors in microwave applications therefore demands even more careful attention to avoiding the presence of such defects in spite of a more complicated means of fabrication.

\begin{figure*}
\includegraphics[width=1\textwidth]{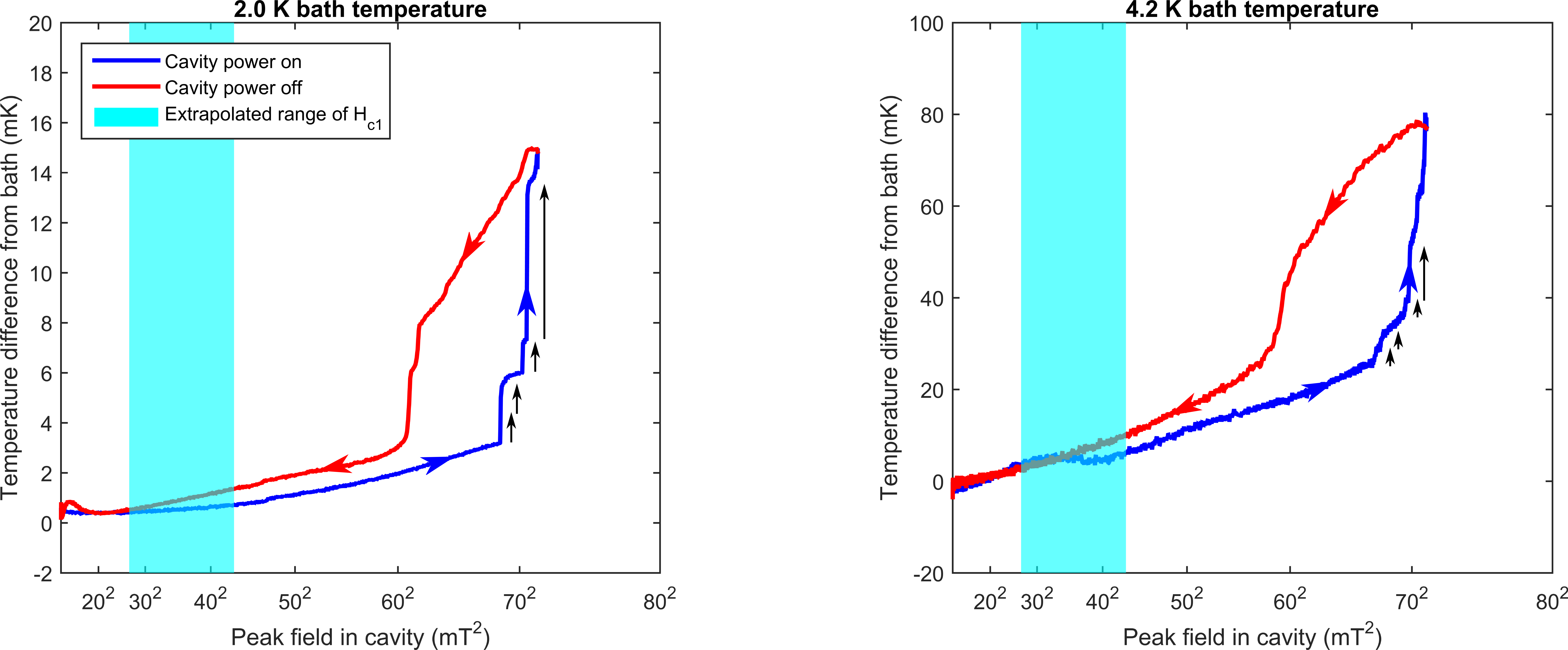}  
\caption{\label{fig:dT_vs_H2} Examples of traces acquired using the sensor at the quench origin, plotted as the temperature rise at the quench origin $\Delta T$ against the square of the local peak surface field $B_{pk}$. The location of the jumps are indicated by arrows. The shaded region indicates the expected range of the lower critical field $B_{c1}$ as extrapolated from material parameters.}
\end{figure*}

\begin{figure}
\includegraphics[width=0.4\textwidth]{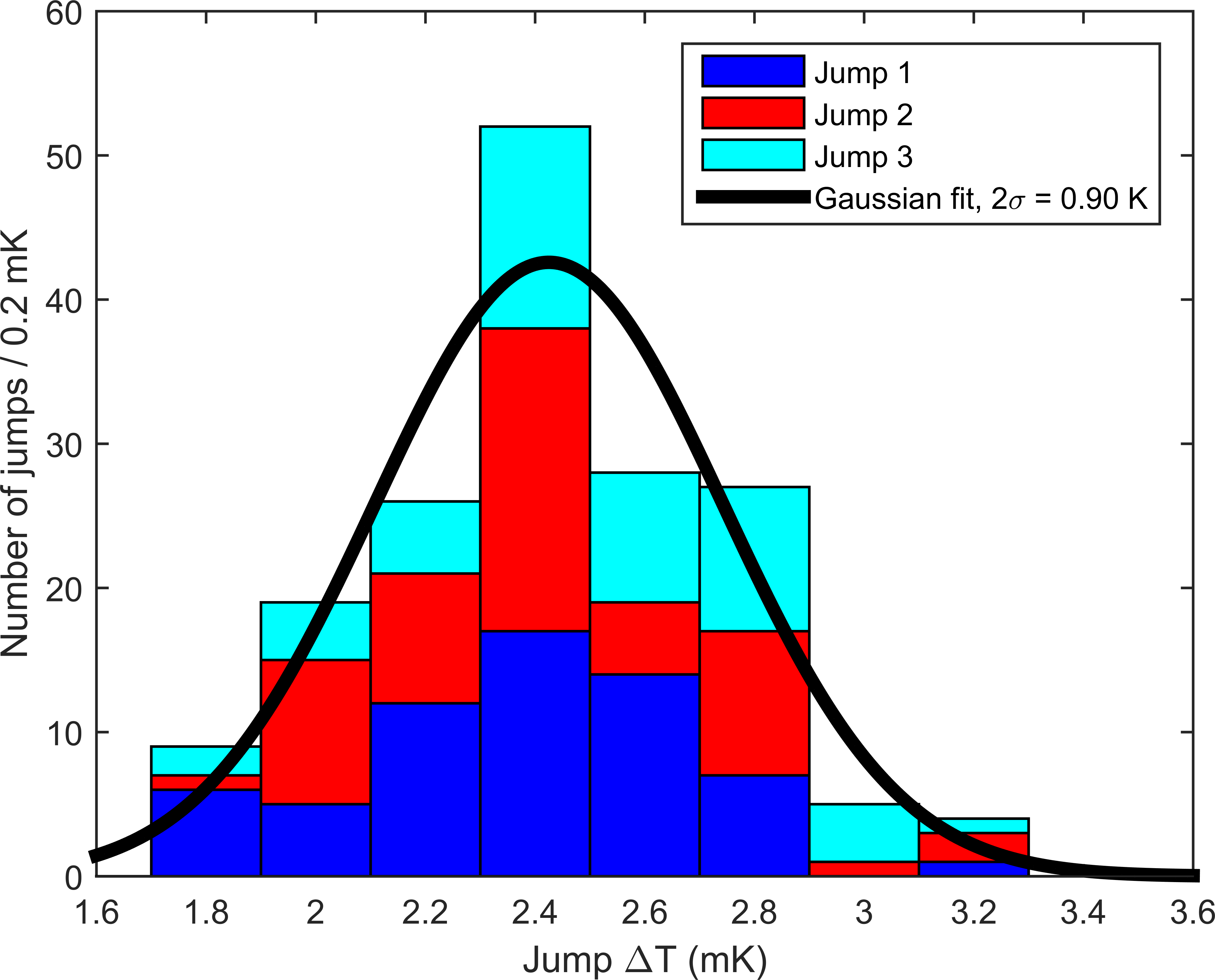}  
\caption{\label{fig:dT_histogram} Stacked histogram of the height of the first three jumps observed in the traces taken at a bath temperature of 4.2~K, indicating that the jumps are of identical height to within the precision of the measurement.}
\end{figure}

In this letter we provide experimental evidence for quantised magnetic vortex entry into a thick polycrystalline film of the Type-II superconductor Nb$_3$Sn in a manner consistent with vortex entry at grain boundaries modelled as underdamped Josephson junctions \cite{PhysRevB.95.214507}. This A15 intermetallic alloy is of particular interest to the particle accelerator community\cite{sustreview}, as its higher transition temperature would allow operation at 4.2~K -- as opposed to the 2.0~K commonly used for current $\approx$1~GHz niobium structures -- and therefore increase the efficiency of the cryogenic plant by up to 80\% over contemporary designs, with an associated reduction in operational costs. Furthermore, modern accelerator specifications demand accelerating gradients of 16-35~MV/m \cite{Bishop2015,Behnke:2013xla}, which in turn requires that the superconducting surface withstand peak surface magnetic fields between 60 and 150~mT, with future demands expected to be even higher. Nb$_3$Sn's superheating field of $\approx$400~mT\cite{Posen2014,Transtrum2011} suggests operation at gradients up to 90~MV/m, an attractive prospect for future projects. However, although Nb$_3$Sn has demonstrated stable operation at fields above the values of $H_{c1}$ \cite{samapl15} extrapolated from material parameters, at time of writing these cavities remain limited to peak surface fields of up to 100~mT \cite{sustreview}, far below $H_{sh}$.


The experiment was carried out using a 1.3~GHz single-cell R\&D niobium cavity of the TeSLA design\cite{samapl15} coated with a layer of Nb$_3$Sn in a custom-built coating furnace\cite{Posen2014,talksrf2015} using the vapour diffusion method\cite{Saur1962}. Although the layer is thin by mechanical standards -- the coating is between 2-3~$\upmu$m thick \cite{ipac2017surface} -- it is many times thicker than the RF penetration depth $\lambda$~nm\cite{Posen2014}, rendering this for all intents and purposes a bulk superconductor. The cavity was operated in a cryostat, immersed in liquid helium, with an antenna mounted to couple in RF power supplied by a solid state amplifier. An RF control system allows operation at a fixed peak RF field in the cavity. This system is used to measure the intrinsic quality factor $Q_0$ and frequency $f$ of the cavity as a function of peak RF field $B_{pk}$ and/or temperature $T$. Measurements of $Q$ vs $T$ and $f$ vs $T$ were analysed with the SRIMP fitting code\cite{halbritter1970comparison,Valles2013,ciovati2004fvsT} to determine the electron mean free path $l = (4.7 \pm 2.0)$~nm. Using literature values\cite{hein1999high} of the London penetration depth $\lambda_L = (89 \pm 9)$~nm and intrinsic coherence length $\xi_0 = 7 \pm 0.7$~nm, we calculate\cite{hein1999high,Tinkham2012} the thermodynamic critical field and lower critical field to be $B_c = (487 \pm 3)$~mT and $B_{c1} = (35 \pm 8)$~mT, respectively.

Information regarding local heating on the inner surface of the cavity during operation is provided by an array of 646 cryogenic temperature sensors affixed to the outside of the cavity, a system dubbed a temperature map or T-map \cite{knoblochthesis}. Use of this temperature mapping system allows for precise local measurements of the heating -- and by extension, power dissipation -- of the superconducting surface as a function of the applied surface RF field. Since previous experiments indicate that the limitation of the field that can be achieved in the cavity is due to a localised defect\cite{ipac2017quench, posenthesis}, as opposed to a global effect, the first stage of the experiment was to use the apparatus to isolate the region at which the superconducting state first broke down when the cavity exceeded its quench field of approximately 75~mT peak surface magnetic field. This was done by quenching the cavity multiple times while using the T-Map to localise the sudden spike in heating (many Kelvin) that accompanies a region being driven into the normal conducting state during quench. 


Once isolated, the quench origin was studied using high-speed, high-resolution operation of a single sensor. The cavity was cycled through $T_c$, to remove any residual trapped magnetic flux introduced by the quench mapping measurement. With the origin of the quench now isolated to a single sensor, data taking began. This consisted of acquiring a set time interval of temperature data from the chosen sensor, during which time the field in the cavity was cycled from empty up to a desired field and then left to ring down. A pickup probe mounted on the end of the cavity allowed precise measurement of the surface magnetic field during the measurement. The result is a combined measurement of the temperature at the origin of the quench minus the temperature of the helium bath, $\Delta T$, and the peak surface magnetic field at this location, $B_{pk}$. This is then recast by equating times and plotting the result as $\Delta T$ against $B_{pk}^2$. The choice of the squared field is due to the fact that conventionally expected Ohmic losses obeying $P \propto B_{pk}^2$ will appear linear when plotted.


Although the cavity was operated very close to the known quench field of the cavity during data-taking, great care was taken in not allowing such a quench to occur. The results are therefore not affected by significant amounts of trapped flux at the quench origin. At fields up to and above $B_{c1}$, the losses appear Ohmic to first order. However, when approaching (yet not exceeding) the quench field, the heating at the quench origin exhibited sudden, discrete jumps, and a hysteresis during the ramp-up in field and ramp-down. This behaviour was seen even after cycling through $T_c$ again, and while operating both at 2.0~K and 4.2~K bath temperatures. Examples of measurement traces exhibiting these jumps, obtained at bath temperatures of 2.0~K and 4.2~K, are shown in Figure \ref{fig:dT_vs_H2}.


These jumps exhibit a number of interesting features. Firstly, during a given cooldown (corresponding to one day of measurements), the field at which any single jump occurs, and the height (in mK) of this jump, is highly repeatable between traces. The second interesting feature is the hysteresis that occurs when these jumps exhibit themselves, with similar sudden drops in temperature visible during the ramp-down measurement. Variations in field and jump height were found to be within the measurement error of extracting these from the trace. In particular, the heights of the first three jumps within any given measurement set were found to be identical within measurement error. This is demonstrated in Figure \ref{fig:dT_histogram}, a stacked histogram of the height of the first three jumps in the measurement data taken at 4.2~K over the course of one day. The stacked distribution is, to first order, Gaussian in nature, with a 2-$\sigma$ interval of 0.9~mK, which is comparable to the precision of the measurement $(\pm 1.2)$~mK for the first three jumps.

\begin{figure}
\includegraphics[width=0.4\textwidth]{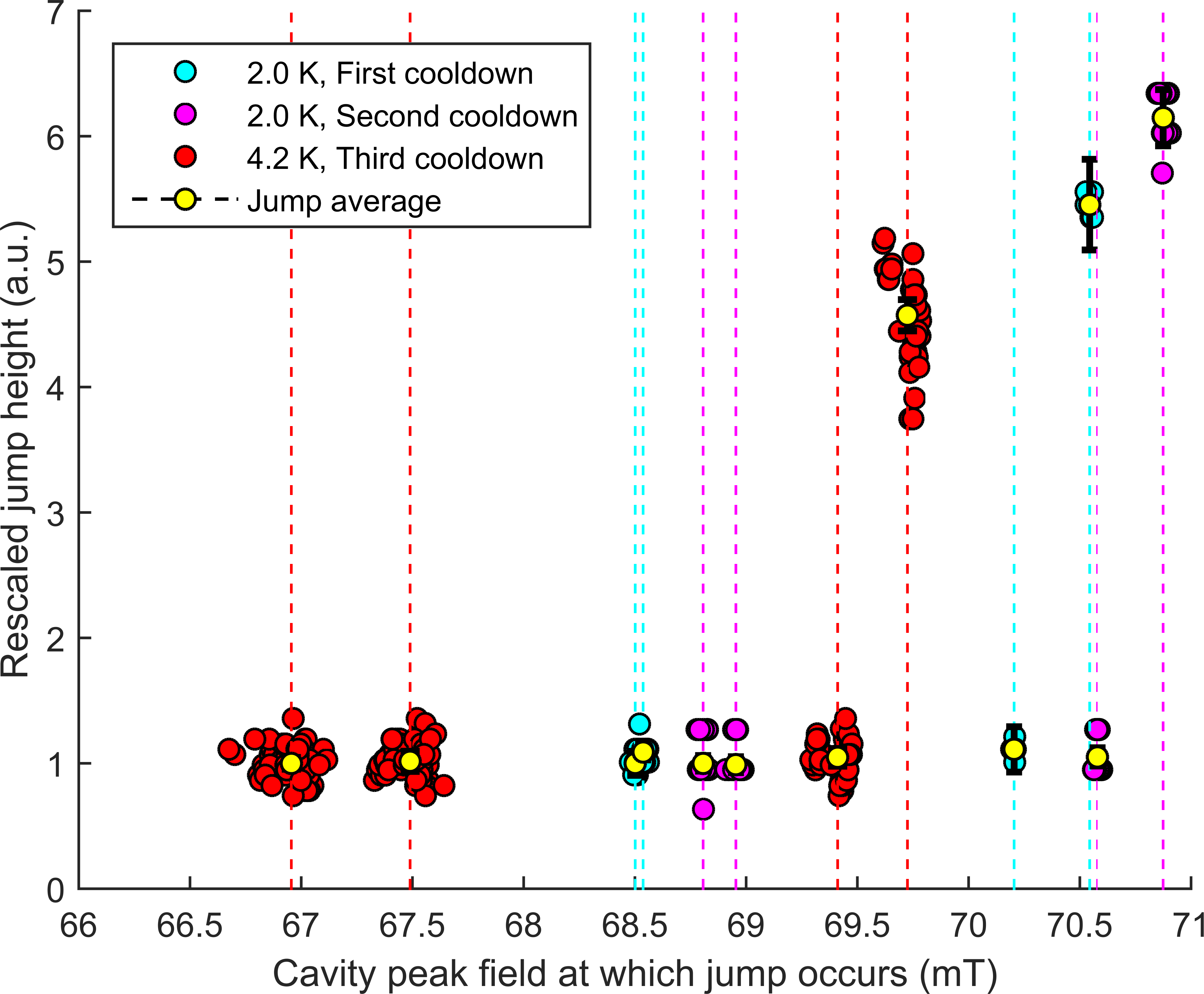}  
\caption{\label{fig:dT_H_scatterplot} Scatter plot of the temperature jumps seen during the three cooldowns, normalised to the mean of the first jump in the trace. The means of the jumps are overlaid.}
\end{figure}

Between cooldown cycles, and at different bath temperatures, the thermal contact between the sensors and the cavity wall varied, changing the absolute (but not relative) measured heights of the jumps day-to-day. This prevented absolute jump height comparison between days; however, normalising all jumps in a cooldown by the mean of the height of the first jump in that cooldown allows for the relative heights to be compared. Rescaling the data from the three separate cooldowns, two at 2.0~K and one at 4.2~K, the jumps are compared in the scatterplot shown in Figure \ref{fig:dT_H_scatterplot}.

\begin{figure}[t]
\includegraphics[width=0.4\textwidth]{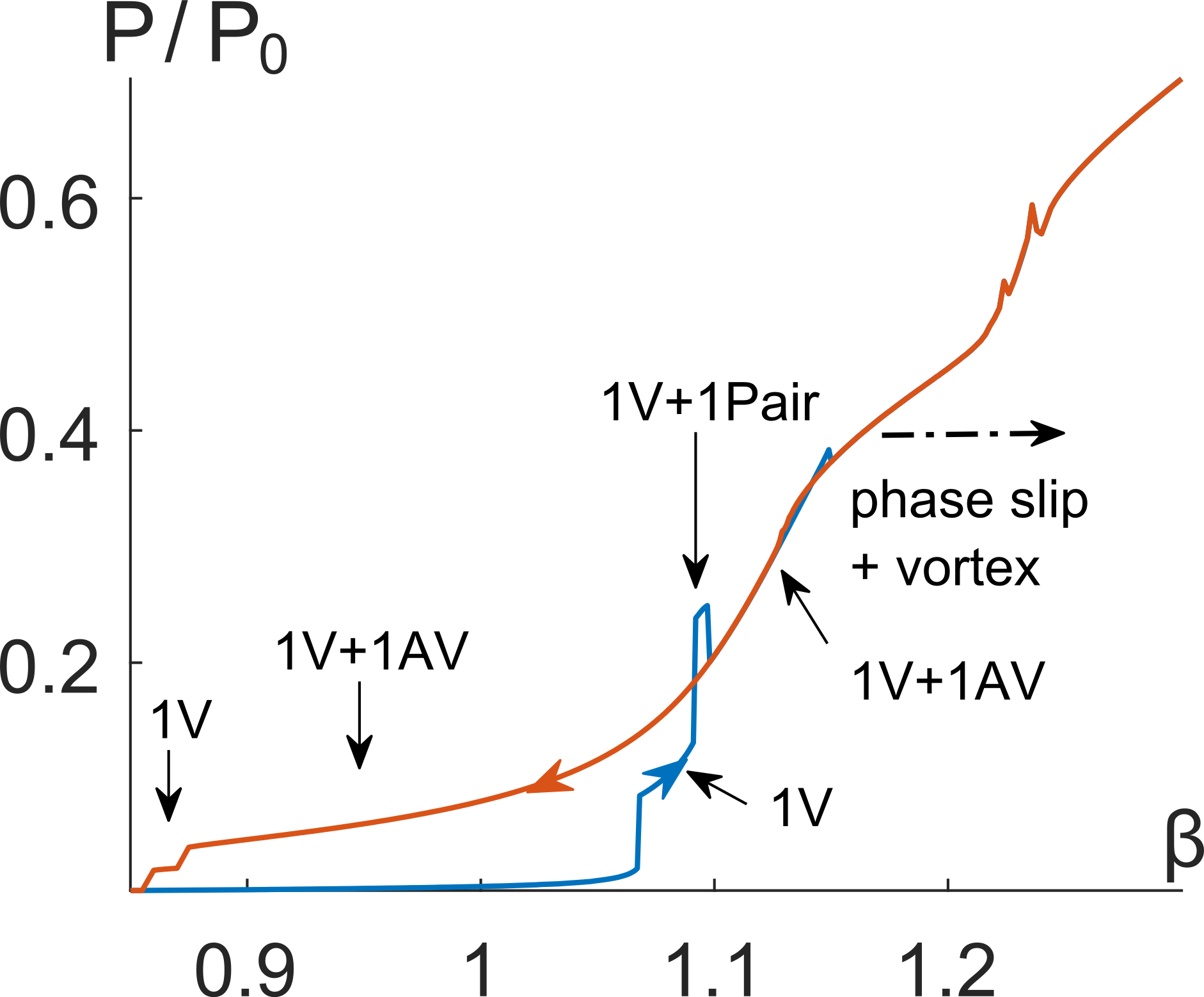}  
\caption{\label{fig:gurevich} Expected behaviour of a grain boundary modelled as an underdamped Josephson junction, plotted as the normalised power dissipated in the grain boundary against the normalised peak AC current through the grain boundary. This figure is reproduced from Ref. \cite{PhysRevB.95.214507} with permission of the author, and is analogous to Figure \ref{fig:dT_vs_H2}. }
\end{figure}


During the first cooldown, measured at 2.0~K, four jumps were seen; $\Delta T = (0.99, \, 1.08, \, 1.10, \, 5.40)$~mK at $B_{pk} = (68.50, \, 68.54, \, 70.21, \, 70.54)$~mT, respectively. Rescaling the first mean to 1, this gives the four jumps as $(1.00 \pm 0.10)$ (forced), $(1.09 \pm 0.07)$, $(1.11 \pm 0.18)$, and $(5.46 \pm 0.36)$. The second cooldown, a repeat of the measurement at 2.0~K, gave $\Delta T = (0.32, \, 0.31, \, 0.33, \, 1.94)$~mK at $B_{pk} = (68.81, \, 68.95, \, 70.58, \, 70.87)$~mT, respectively. The decrease in the jump height can be attributed to a decrease in the thermal contact between the temperatures sensors and the cavity outer wall. Again, rescaling the first jump to 1 gives $(1.00 \pm 0.06)$ (forced), $(0.99 \pm 0.07)$, $(1.05 \pm 0.08)$, and $(6.15 \pm 0.23)$. The third cooldown, a measurement performed at 4.2~K, found $\Delta T = (2.43, \, 2.47, \, 2.54, \, 11.11)$~mK at $B_{pk} = (66.96, \, 67.49, \, 69.41, \, 69.73)$~mT, respectively. Rescaling gives $(1.00 \pm 0.05)$ (forced), $(1.02 \pm 0.08)$, $(1.05 \pm 0.07)$, and $(4.57 \pm 0.13)$. Within any one cooldown, each independent jump is reproducible in both the magnetic field at which it occurs and the height of the jump, to within the precision of the measurement. However, between cooldowns, drifts in the cable calibration used to extrapolate the surface field result in the introduction of a systematic error. The difference between the fields at which the jumps occur is on the order of this error and therefore differences observed between cooldowns are not deemed significant.

A mechanism that could explain this behaviour is quantised flux entry at grain boundaries modelled as Josephson junctions, as proposed by Sheikhzada and Gurevich\cite{gurevich2017a,PhysRevB.95.214507}. In this scenario, a grain boundary whose thickness is on the order of the bulk coherence length acts as a blocking defect for the supercurrent. Above a grain boundary-specific critical field $B_{GB}$, magnetic flux enters in a quantised manner. These vortices are not expected to be of pure Abrikosov nature, as is usually the case with magnetic flux pinned during the superconducting transition, but of a mixed Abrikosov-Josephson vortex with a phase core of length\cite{gurevich2017a} $l = \lambda_J^2 / \lambda$ where $\lambda_J \sim \sqrt{\xi \lambda B_c / B_{GB}}$. The speed of a vortex along a grain boundary is expected\cite{Embon2017} to be on the order of 10~km/s, and so the time of passage of a vortex along a grain boundary of 3~$\upmu$m is on the order of half the RF period of 0.76~ns. As the vortex travels along the grain boundary, it dissipates energy due to a viscous drag $\eta = 1/(\omega_J R C)$, where $R$ is the resistance of the grain boundary, $C$ its specific capacitance, and $\omega_J$ is the Josephson frequency\cite{PhysRevB.95.214507}.

An example of the power dissipated in a grain boundary modelled as an underdamped ($\eta < 1$) junction as a function of the current across the boundary is shown in Fig. \ref{fig:gurevich}, provided by Sheikhzada and Gurevich. This figure is analogous to the measurement shown in Fig. \ref{fig:dT_vs_H2}, and both demonstrate similar features of sudden jumps in temperature/dissipated power and hysteresis with applied current/RF field. Provided the mechanisms are the same, the similar height of the first three temperature jumps shown in Fig. \ref{fig:dT_H_scatterplot} would then be explained by the quantised entry of three single vortices (one additional vortex per jump) into a grain boundary. Although the power dissipation by a vortex is a function of the applied RF field, the difference in the applied RF field between the four jumps is smaller than the precision of the measurement ($\pm 4$\% against $\pm 6$\%) and so any difference is expected to be undetectable. The hysteresis is explained by the fact that once vortices begin to enter the grain boundary, they remain inside until the RF field is reduced to a value below that at which they originally entered. They can even remain for longer if pinned or slowed by defects in the grain boundary, transitioning to anti-vortices as the RF completes enters the second half of the cycle. As the applied RF field increases, the behaviour becomes more complicated as more vortices enter, and eventually a phase slip will occur that will drive the entire grain boundary into the normal conducting state. The expectation is that once this occurs, the dissipation at the grain boundary will be so great as to drive the local region into the normal conducting state, and the rest of the cavity follows shortly thereafter.

Once a grain boundary is in the phase slip state, the power dissipated approaches an ohmic dependence of $P \approx R \mu_0^2 ( B_{pk}^2 - B_{GB}^2 ) / (2 \lambda^2)$ \cite{gurevich2017a}. Given that the grain boundary critical field is expected to lie at the centre of the hysteresis seen in Fig. \ref{fig:dT_vs_H2}, then $B_{GB} \approx 65$~mT. Assuming a grain boundary thickness of 2~nm, a penetration depth of 140~nm, and a normal conducting resistivity of 1~$\upmu\Omega$m, then a grain boundary in the phase slip state 1~micron wide by 1~micron deep dissipates $\sim 50$~$\upmu$W at the quench field of $\approx 75$~mT. A larger grain boundary, 3~micron wide and 3~micron deep (extending through the entire thickness of the layer), would dissipate  $\sim 400$~$\upmu$W. These surface field values assume no surface field enhancement due to topology, which can be as high as a factor of 2, which would raise these values to $180$~$\upmu$W - 1.6~mW. Taking into consideration a realistic range of values of the penetration depth, grain boundary size, and field enhancement factor, the expected range for the dissipation at the quench field is $30$~$\upmu$W - 3~mW.

By calibrating the $\Delta T$ to the dissipated power in the cavity, we find that the power dissipated at the quench origin just before the quench field is $(2.2 \pm 0.5)$~mW, which is qualitatively consistent with the calculation above. Furthermore, numerical simulations such as those shown in Fig. \ref{fig:gurevich} for an underdamped grain boundary suggest that the increase in power dissipation from a single vortex should be of order $1/10$ the phase slip value, which is again consistent with the first three jumps each corresponding to an increase in dissipated power of $(130 \pm 30)$~$\upmu$W.

To conclude, evidence for quantised vortex entry in Nb$_3$Sn under the influence of an RF field has been presented. The behaviour observed is consistent with that expected from the entry of mixed Abrikosov-Josephson vortices into a grain boundary acting as a strongly-coupled, under-damped Josephson junction with $B_{GB} \approx 0.1 B_{c}$. This result would explain the quench observed in Nb$_3$Sn microwave cavities as due to the surface being incapable of sustaining the dramatically increased losses generated by a grain boundary being driven into the phase slip state, and suggests that in order to increase the quench field of these cavities the grain boundaries must be made even more strongly coupled to shift $B_{GB} \rightarrow B_c$. This could be achieved by increasing the coherence length $\xi$ by increasing the bulk electron mean free path, ``cleaning'' the superconductor. Although reducing the mean free path in a controlled manner has been well documented in conventional niobium cavities using impurity doping\cite{Grassellino2013,maniscalco2017mfp}, there is not yet an established means of increasing it that is appropriate for Nb$_3$Sn coatings. Future work on Nb$_3$Sn in microwave cavities will focus on altering such material parameters in a controlled manner to tailor the performance of the device. We hope that the outcome of this effort will lead to a better understanding of the behaviour of strongly Type-II, polycrystalline superconductors in microwave applications, and to the development of more efficient and powerful superconducting accelerator cavities for application in science and industry.

This work was supported by U.S. Department of Energy (DOE) award DE-SC0008431. The authors would like to acknowledge and thank Prof. A.V. Gurevich at Old Dominion University and Prof. J.P. Sethna and Dr. D.B. Liarte at Cornell University for their helpful contributions and discussions.


%
%

%


\bibliography{bibfor_evidence_for_quantized_vortex}

\end{document}